\documentclass[twocolumn,showpacs,amsmath,amssymb,superscriptaddress,aps,pra]{revtex4-1}
\usepackage{amssymb,amsmath}
\usepackage{graphicx}
\usepackage{enumerate}
\usepackage{paralist}
\usepackage{dsfont}
\usepackage{color}

\newcommand{\ket}[1]{| #1 \rangle}
\newcommand{\bra}[1]{\langle #1 |}

\begin{document}

\title{Nonadiabatic quantum state engineering driven by fast quench dynamics}
\author{Marcela Herrera}
\email{alba.trujillo@ufabc.edu.br}
\affiliation{Centro de Ci\^{e}ncias Naturais e Humanas, Universidade Federal do ABC, Rua Santa Ad\'{e}lia 166, 09210-170 Santo Andr\'{e}, S\~{a}o Paulo, Brazil}

\author{Marcelo S. Sarandy}
\email{msarandy@if.uff.br}
\affiliation{Instituto de F\'{i}sica, Universidade Federal Fluminense, Avenida Gal. Milton Tavares de Souza s/n, Gragoat\'{a}, 24210-346 Niter\'{o}i, Rio de Janeiro, Brazil}

\author{Eduardo I. Duzzioni}
\email{duzzioni@gmail.com}
\affiliation{Departamento de F\'{i}sica, Universidade Federal de Santa Catarina, Caixa Postal 476, 88040-970 Florian\'{o}polis, Santa Catarina, Brazil}

\author{Roberto M. Serra}
\email{serra@ufabc.edu.br}
\affiliation{Centro de Ci\^{e}ncias Naturais e Humanas, Universidade Federal do ABC, Rua Santa Ad\'{e}lia 166, 09210-170 Santo Andr\'{e}, S\~{a}o Paulo, Brazil}

\date{\today}
\begin{abstract}
There are a number of tasks in quantum information science that exploit non-transitional adiabatic dynamics. Such a dynamics is bounded by the adiabatic theorem, which naturally imposes a speed limit in the evolution of quantum systems. Here, we investigate an approach for quantum state engineering exploiting a shortcut to the adiabatic evolution, which is based on rapid quenches in a continuous-time Hamiltonian evolution. In particular, this procedure is able to provide state preparation faster than the adiabatic brachistochrone. Remarkably, the evolution time in this approach is shown to be ultimately limited by its ``thermodynamical cost,'' provided in terms of the average work rate (average power) of the quench process. We illustrate this result in a scenario that can be experimentally implemented in a nuclear magnetic resonance setup.    
\end{abstract}

\pacs{03.67.-a, 02.30.Zz, 03.65.Ud, 75.10.Dg}

\maketitle

\section{Introduction}

Quantum state engineering (QSE), which aims at manipulating a quantum system to attain a precise target state, is a 
fundamental step in quantum information tasks~\cite{nielsen,suter,qeng}. In this context, the adiabatic theorem of quantum 
mechanics~\cite{messiah} has been consolidated as a valuable tool, allowing for QSE protocols designed to a wide range of 
applications, such as state transfer~\cite{StateTransfer}, dynamics of quantum critical phenomena~\cite{Dyn-QPT}, and 
quantum computation~\cite{Zanardi:99,farhi1,Bacon}. Moreover, adiabatic QSE has been experimentally realized 
through a number of techniques, such as nuclear magnetic resonance (NMR)~\cite{nmr1,nmr2,nmr3}, superconducting qubits~\cite{supercon3},
 trapped ions~\cite{trap1,trap2}, and optical lattices~\cite{opt}.  On general grounds, adiabatic QSE is obtained through a 
slowly varying time-dependent Hamiltonian $H(t)$, which drives an instantaneous eigenstate $|n(0)\rangle$ of the initial 
Hamiltonian $H(0)$ to the corresponding instantaneous eigenstate $|n(T_{ad})\rangle$ of the final Hamiltonian $H(T_{ad})$. 
The success of the protocol can be  estimated by a distance measure (such as the trace distance~\cite{nielsen})  between 
the final and target states. This performance  turns out to be related to the speed of the evolution, which is upper bounded by 
the gap structure of the energy spectrum of $H(t)$ (see, e.g., Ref.~\cite{sarandy2004}).  

While the adiabatic passage provides a faithful procedure to achieve a designed target state in isolated systems, 
it may undergo a considerable fidelity loss in the presence of system-bath interactions, which induces decohering 
effects in the quantum evolution. Indeed, for systems under decoherence, there is a competition between the time 
required for adiabaticity and the decoherence time scales~\cite{Sarandy:05}, which typically limits the success of 
the adiabatic approach. Recently, as an alternative direction, several nonadiabatic protocols related to QSE have 
been proposed~\cite{Berry:09,Chen:11,sarandy2011,utkan}. Berry~\cite{Berry:09} introduced an approach 
named {\it transitionless quantum driving}, in which Hamiltonians are designed in order to follow a path similar to 
the adiabatic one in an arbitrary time. In Ref.~\cite{Chen:11} relations between the Berry and the dynamic invariant 
approaches are discussed in terms of an explicit example of a nonadiabatic shortcut for harmonic trap expansion and 
state preparation of two-level systems. The dynamic invariant approach was also employed to gather a shortcut to 
adiabatic quantum algorithms, such as in the Grover and Deutsch-Jozsa problems~\cite{ sarandy2011}. In~\cite{utkan}, 
a method based on Lie algebras is presented to build dynamic invariants for four-level-system Hamiltonians. Transitionless 
quantum driving was also studied for the scenario of open-system dynamics in Refs.~\cite{Vacanti2013,Jing:2013}, 
where both Markovian and non-Markovian baths are considered. As a further contribution to these previous results, 
we introduce here a nonadiabatic QSE protocol based on a fast-quench dynamics governed by a Hamiltonian, which is 
experimentally realizable in an NMR setup. Moreover, we analyze resources involved in nonadiabatic QSE from a 
thermodynamical perspective, concluding that the shortcut to the adiabatic case is limited by the energetic cost of the 
evolution, provided in terms of the average work rate (average power) of the quench process.

The starting point to our implementation of QSE by fast quenches is the inverse-engineering approach~\cite{Chen:11,sarandy2011}, 
which has been used in a number of quantum control applications (see Ref.~\cite{Muga:Review} for a recent review). 
More specifically, instead of considering an interpolating time-dependent Hamiltonian from the beginning, we will take the 
evolution of an initial quantum state $|\psi(t_0)\rangle$ by means of a Lewis-Riesenfeld dynamic invariant 
$\mathcal{I}(t)$~\cite{lewis1,lewis2}, which is a Hermitian operator capable of providing the evolved state $|\psi(t)\rangle$ 
at any time $t>t_0$. The dynamic invariant is built up in such a way as to connect the evolution of the initial state to the target 
desired state. Then,  from the definition of $\mathcal{I}(t)$, we can inversely derive a physical Hamiltonian that effectively 
drives the system to the target state at a final time $T$ in a nonadiabatic regime. We will employ this method to define a 
quench dynamics that yields a shortcut to the best adiabatic QSE protocol. 
Remarkably, the speed of evolution in the nonadiabatic protocol will be shown to be upper bounded by the average work 
associated with the implementation of the quench process. 

\section{Lewis-Riesenfeld invariants}  
\label{continuous}

We start by briefly reviewing the Lewis-Riesenfeld dynamic invariant formalism.  For a closed quantum system described 
by a time-dependent Hamiltonian $H(t)$, a dynamic invariant $\mathcal{I}(t)$ is defined as a Hermitian operator that 
satisfies~\cite{lewis1,lewis2}
\begin{equation}
\frac{\partial \mathcal{I}(t)}{\partial t}+i\left[H(t),\mathcal{I}(t)\right]=0,
\label{eqinv}
\end{equation}
where $\hbar$ has been set to 1. From the instantaneous spectral decomposition of $\mathcal{I}(t)$, we write
\begin{equation}
\mathcal{I}(t)=\sum_j \lambda_j |\phi_j (t)\rangle\langle\phi_j (t)|,
\end{equation}
where $\{\lambda_j\}$ denotes the eigenvalue set of $\mathcal{I}(t)$, which is associated with the set of eigenstates 
$\{|\phi_j (t)\rangle\}$. For simplicity, we will assume that the eigenvalue spectrum of $\mathcal{I}(t)$ is nondegenerate.  
Moreover, Eq.~(\ref{eqinv}) implies that the eigenvalues $\lambda_j$  are real and time independent. Equivalently, it 
follows that the dynamic invariant is a conserved quantity, i.e., $d \langle \mathcal{I}(t) \rangle / dt = 0$. The dynamic 
invariant $\mathcal{I}(t)$ can be used to determine the general solution of the Schr\"{o}dinger equation 
\begin{equation}
i \frac{d}{dt}|\psi(t)\rangle=H(t)|\psi(t)\rangle.
\label{td-se}
\end{equation}
More specifically, the state vector $|\psi(t)\rangle$ in Eq.~(\ref{td-se}) can be expressed as the linear superposition of 
the instantaneous eigenstates of the invariant Hermitian operator \cite{lewis2},
\begin{equation}
|\psi(t)\rangle=\sum_j c_j(t)|\phi_j(t)\rangle,
\end{equation}
where $c_j(t)$ satisfies
\begin{equation}
c_{j}(t)=c_{j}(0)e^{\left[  -\int_{0}^{t}d\tau\left(  \langle\phi_{j}|\frac{\partial}{\partial\tau}|\phi_{j}\rangle+
i\langle\phi_{j}|H|\phi_{j}\rangle\right)  \right]  }.
\end{equation}
Therefore, if we initially prepare the system in the $j$th eigenstate $|\phi_{j}(0)\rangle$ of $\mathcal{I}(0)$ then the 
system will necessarily evolve to $|\phi_{j}(t)\rangle$ at any time $t$, since the the dynamic invariant is a conserved quantity. 
This result can be employed to tailor an evolution towards an arbitrary target state. To this aim, let us consider the 
parametrized (normalized) time $\tau=t/T$, where $T$ denotes the final time in the system evolution.  Then the dynamic 
invariant $\mathcal{I}(\tau)$ can be defined in a such way that (a) $\mathcal{I}(0)$ has a nondegenerate eigenstate 
$|\phi(0)\rangle$ (the initial state); (b) $\mathcal{I}(1)$ has a nondegenerate eigenstate $|\phi(1)\rangle$ (the target state); 
and (c) $\mathcal{I}(\tau)$ is obtained, for intermediary values of $\tau$ $(0 < \tau < 1)$, by a conveniently chosen interpolation. 

The conserved quantity $\mathcal{I}(\tau)$ has an associated Hamiltonian that guides the system dynamics, which can be 
determined following the strategy described in Ref. \cite{sarandy2011}, where the invariant and the Hamiltonian are written 
as a linear combination of the generators of a Lie algebra, with time-dependent coefficients to be determined by Eq.~(\ref{eqinv}) 
and by the initial conditions. In order to connect $\mathcal{I}(\tau)$ with $H(\tau)$, we apply a change of variables in 
Eq.~(\ref{eqinv}), replacing $t$ with $\tau$. This yields
\begin{equation}
\frac{\partial \mathcal{I}(\tau)}{\partial \tau}+Ti \left[H(\tau),\mathcal{I}(\tau)\right]=0 .
\label{eqinv2}
\end{equation}
From the solution of Eq.~(\ref{eqinv2}) (with fixed initial conditions), we can completely determine $H(\tau)$. 
To test the effectiveness of this approach, we compare the nonadiabatic route with the best adiabatic evolution 
employing the trace distance $\delta(\tau)$ between the evolved state $\rho(\tau)$ and the target state $\rho(\tau=1)$, which reads

\begin{equation}
\delta(\tau) \equiv \frac{1}{2}\text{tr}\sqrt{\left[\rho(\tau)-\rho(\tau=1)\right]^2}.
\label{tracedistance}
\end{equation}
Observe that $\delta(\tau)$ is directly related to the fidelity of the QSE protocol, with $\delta(\tau)=0$ implying perfect 
preparation of the target state.
%
\section{Preparation of entangled states}
\label{entangled}

In order to introduce the nonadiabatic QSE protocol by fast quenches, we will consider the evolution of a separable state 
$|\psi_{0}\rangle$ to a maximally entangled state $|\psi_{f}\rangle$ (Bell state), with 
\begin{eqnarray}
|\psi_{0}\rangle&=&|j,k\rangle , \nonumber \\
|\psi_{f}\rangle&=&\frac{1}{\sqrt{2}}\left[|j,k\rangle+(-1)^{k}|j\oplus 1,k\oplus 1\rangle\right] , 
\label{QSE-ex}
\end{eqnarray}
with $j,k=0,1$. Even though we will focus on a specific example of QSE, the method is general, being applicable 
for a general state preparation. For reasons of comparison, we will start by designing an adiabatic procedure for the  
generation of $|\psi_{f}\rangle$ based on the quantum adiabatic brachistochrone (QAB)~\cite{zanardi}. Then we will proceed by 
proposing a nonadiabatic dynamics through a suitable choice of a class of dynamic invariants, which in turn allows 
for a benchmark analysis of performance with respect to the adiabatic protocol.

\subsection{Adiabatic QSE}
\label{aqse}

The continuous time preparation of the target state $|\psi_{f}\rangle$ from the initial state $|\psi_{0}\rangle$ can be obtained 
by an adiabatic interpolation between  the initial Hamiltonian $H_{0}=\omega(\mathds{1}-|\psi_{0}\rangle\langle\psi_{0}]|)$ 
and the final Hamiltonian $H_{f}=\omega(\mathds{1}-|\psi_{f}\rangle\langle\psi_{f}|))$, where $\omega$ sets 
an energy scale. This will be implemented through a time-dependent 
Hamiltonian 
\begin{equation}
H_A[s(t)] = [1 -  s(t)]H_{0} + s(t)H_{f}, 
\label{adiab-inter}
\end{equation}
with $s(t)$ satisfying the local adiabatic condition~\cite{roland} 
\begin{equation}
\left| \frac{ds}{dt} \right| \le \varepsilon \frac{\Omega^2(s)}{\langle d H_A / ds \rangle_{1,0}},
\label{local-adiab}
\end{equation}
where $\varepsilon<1$ is the probability of finding the system in an excited state, $\Omega(s)$ is the gap between 
the two lowest eigenvalues, and $\langle d H_A / ds \rangle_{1,0}$ is the matrix element of 
$d H_A / ds $ between the two corresponding eigenstates. In our case, we will be interested in the interpolation 
that employs the shortest time evolution that satisfies the adiabatic condition. For this purpose, we will focus on 
the geometric approach introduced in Ref.~\cite{zanardi}. The optimal time for the adiabatic quantum evolution 
is associated with the optimal trajectory followed by the quantum system in Hilbert space, which is defined as the QAB. 
The problem of the QAB is time-locally equivalent to finding the shortest path between two points in a complex 
space~\cite{zanardi}.  For the interpolation given by Eq.~(\ref{adiab-inter}), the best interpolating function 
$s_{AQB}(\tau)$ is then given by~\cite{zanardi}
\begin{equation}
s_{QAB}(\tau)=\frac{1}{2}-\frac{\left|\alpha_{0}\right|}{2\sqrt{1-\left|\alpha_{0}\right|^{2}}}\tan\left[\left(1-2\tau\right)
\arccos\left|\alpha_{0}\right|\right],
\label{xbaq}
\end{equation}
where $\tau = t/T_{ad} \, \in\left[0,1\right]$ and $T_{ad}$ is the total time for the adiabatic evolution. From Eq.~(\ref{QSE-ex}), 
we obtain that $\left|\alpha_{0}\right|=\left|\langle\psi_{0}|\psi_{f}\rangle\right|=1/\sqrt{2}$.  
Then, by inverting Eq.~(\ref{xbaq}), we can rewrite it as 
\begin{equation}
\tau = \frac{t}{T_{ad}} =\frac{2}{\pi} \left\{ \frac{\pi}{4} + \arctan \left[\left(2s-1\right)\right]\right\}.
\label{xbaq2}
\end{equation}
The total adiabatic time $T_{ad}$ can be derived from Eq.~(\ref{local-adiab}) (see, e.g., Ref.~\cite{roland}), yielding here 
\begin{equation}
T_{ad}=\frac{\pi}{2\epsilon\omega}. 
\label{Tad}
\end{equation}
Note that $T_{ad}$ is inversely proportional to the energy scale $\omega$. 
Furthermore, it is also inversely proportional to $\varepsilon$ (the fail probability); thus the adiabatic time should be large enough 
to reach the final state with high probability. 

\subsection{Nonadiabatic QSE}
\label{naqse}

In order to explore a nonadiabatic mechanism to drive the system continuously in time to the desired target entangled state, 
we consider a class of dynamic invariants built up in terms of two spin-1/2 operators arranged at sites $j=1,2$, which is given by
\begin{eqnarray}
\mathcal{I}(\tau)&=&G_{1}(\tau)\sigma_{z}^{(1)}+G_{2}(\tau)\sigma_{z}^{(2)}+G_{3}(\tau)\sigma_{y}^{(1)}\sigma_{x}^{(2)}\nonumber\\
&+&G_{4}(\tau)\sigma_{x}^{(1)}\sigma_{y}^{(2)}+G_{5}(\tau)\sigma_{x}^{(1)}\sigma_{x}^{(2)}+G_{6}(\tau)\sigma_{y}^{(1)}\sigma_{y}^{(2)}, \nonumber \\
\label{di-su2}
\end{eqnarray}
where $G_{j}(\tau)$ denote the coefficients of the dynamic invariant and $\sigma_{k}^{(j)}$ are the Pauli operators at site 
$j$ (for $k=x,y,z$). The motivation for the proposal of $\mathcal{I}(\tau)$ as in Eq.~(\ref{di-su2}) is that it ensures a dynamics 
governed by a two-spin Hamiltonian of the form
\begin{equation}
H_{I}(\tau)=J\pi\sigma_{x}^{(1)}\sigma_{x}^{(2)}+f(\tau)(\sigma_{z}^{(1)}+\sigma_{z}^{(2)}), 
\end{equation}
where $J$ is the scalar coupling and $f(\tau)$  is a time-dependent modulation function. This is the typical Hamiltonian of 
two coupled spins used in a scalar molecule in an NMR system \cite{oliveira, maziero, oliveira2}. To connect $\mathcal{I}(\tau)$ and $H_{I}(\tau)$, 
we are  required to solve a set of coupled differential equations arising from Eq.~(\ref{eqinv2}). This problem is simplified through a 
convenient change of variables:
\begin{eqnarray}
G_{1}(\tau)&=&\frac{g_{1}(\tau)+g_{4}(\tau)}{2}, \,\,\,\,\, G_{2}(\tau)=\frac{g_{1}(\tau)-g_{4}(\tau)}{2}, \nonumber \\
G_{3}(\tau)&=&\frac{g_{2}(\tau)+g_{5}(\tau)}{2}, \,\,\,\,\, G_{4}(\tau)=\frac{g_{2}(\tau)-g_{5}(\tau)}{2},\,\,\,\,\, \\
G_{5}(\tau)&=&\frac{g_{3}(\tau)+g_{6}(\tau)}{2},\,\,\,\,\, G_{6}(\tau)=\frac{g_{3}(\tau)-g_{6}(\tau)}{2}. \nonumber
\end{eqnarray}
In terms of this new set of parameters, we can rewrite $\mathcal{I}(\tau)$ as 
\begin{eqnarray}
\mathcal{I}(\tau)&=&g_{1}(\tau) \Sigma_1^{(1)} - g_{2}(\tau) \Sigma_2^{(1)}+  g_{6}(\tau) \Sigma_3^{(1)} \nonumber\\
&+&g_{3}(\tau) \Sigma_1^{(2)} +  g_{4}(\tau) \Sigma_2^{(2)} -  g_{5}(\tau) \Sigma_3^{(2)},\,\,\,\,\, 
\label{di-g-su2}
\end{eqnarray}
where 
\begin{eqnarray}
 \Sigma_1^{(1)} &=& \left( \frac{\sigma_{z}^{(1)}+\sigma_{z}^{(2)}}{2}\right) , \,\,\,\,\,
 \Sigma_2^{(1)} = -\left( \frac{\sigma_{y}^{(1)}\sigma_{x}^{(2)}+\sigma_{x}^{(1)}\sigma_{y}^{(2)}}{2}\right),\nonumber\\
 \Sigma_3^{(1)} &=& \left( \frac{\sigma_{x}^{(1)}\sigma_{x}^{(2)}-\sigma_{y}^{(1)}\sigma_{y}^{(2)}}{2}\right) 
\end{eqnarray}
 and 
 \begin{eqnarray}
 \Sigma_1^{(2)}&=& \left( \frac{\sigma_{x}^{(1)}\sigma_{x}^{(2)}+\sigma_{y}^{(1)}\sigma_{y}^{(2)}}{2}\right)  , \,\,\,\,\,
 \Sigma_2^{(2)}=\left( \frac{\sigma_{z}^{(1)}-\sigma_{z}^{(2)}}{2}\right) , \nonumber \\  
 \Sigma_3^{(2)}&=&-\left( \frac{\sigma_{y}^{(1)}\sigma_{x}^{(2)}-\sigma_{x}^{(1)}\sigma_{y}^{(2)}}{2}\right)  . 
 \end{eqnarray}
The sets $\{\Sigma_1^{(1)},\Sigma_2^{(1)},\Sigma_3^{(1)}\}$ and  $\{\Sigma_1^{(2)},\Sigma_2^{(2)},\Sigma_3^{(2)}\}$ provide 
two independent su(2) algebras, where $[ \Sigma_i^{(\alpha)},\Sigma_j^{(\alpha)}] = 2 i \varepsilon_{ijk} \Sigma_k^{(\alpha)}$, 
with $\alpha \in \{1,2\}$ and $\varepsilon_{ijk}$ denoting the Levi-Civita symbol [it is $1$ if $(i,j,k)$ is an even permutation of 
$(1,2,3)$, $-1$ if it is an odd permutation, and 0 if any index is repeated]. As also noticed in Ref.~\cite{utkan}, this algebraic 
independent structure of the generators of $\mathcal{I}(\tau)$ allows for the decoupling of Eq.~(\ref{eqinv2}) into two 
independent subsets of differential equations for the coefficients $\{g_i (\tau)\}$, reading 
\begin{subequations}
\label{gequations}
\begin{align}
\dot{g}_{1}(\tau) & =  2\pi TJg_{2}(\tau), \label{g1}\\
\dot{g}_{2}(\tau) & =  4Tf(\tau)g_{6}(\tau)-2\pi TJg_{1}(\tau),\label{g2}\\
\dot{g}_{3}(\tau) & =  0, \label{g3}\\
\dot{g}_{4}(\tau) & =  2\pi TJg_{5}(\tau),\label{g4}\\
\dot{g}_{5}(\tau) & =  -2\pi TJg_{4}(\tau), \label{g5}\\
\dot{g}_{6}(\tau) & =  -4 Tf(\tau)g_{2}(\tau) \label{g6} .
\end{align}
\end{subequations}
It is worthwhile to mention that only the subset $\{g_1 (\tau),g_2(\tau),g_6 (\tau)\}$, governed by Eqs.~\eqref{g1}, \eqref{g2}, 
and \eqref{g6}, is  responsible for the system dynamics, involving the time-dependent coefficient $f(\tau)$ of the Hamiltonian. 
As expected, $\mathcal{I}(\tau)$ exhibits time-independent (constant) eigenvalues
\begin{subequations}
\begin{align}
\lambda_{1}&=-\sqrt{g_{3}^{2}(\tau)+g_{4}^{2}(\tau)+g_{5}^{2}(\tau)},\\
\lambda_{2}&=\sqrt{g_{3}^{2}(\tau)+g_{4}^{2}(\tau)+g_{5}^{2}(\tau)},\\
\lambda_{3}&=-\sqrt{g_{1}^{2}(\tau)+g_{2}^{2}(\tau)+g_{6}^{2}(\tau)},\\
\lambda_{4}&=\sqrt{g_{1}^{2}(\tau)+g_{2}^{2}(\tau)+g_{6}^{2}(\tau)},\label{ll4}
\end{align}
\end{subequations}
with associated eigenstates (in the computational basis $\{\left|00\right\rangle,\left|01\right\rangle,\left|10\right\rangle,\left|11\right\rangle\}$) given by
\begin{subequations}
\begin{align}
\small
|\phi_{1}(\tau)\rangle&=\eta^{(a)}_{1}\left(0,\frac{g_{4}(\tau)+\lambda_{1}}{g_{3}(\tau)+ig_{5}(\tau)},1,0\right)^{\text{T}},\label{phi1}\\
|\phi_{2}(\tau)\rangle&=\eta^{(a)}_{2}\left(0,\frac{g_{4}(\tau)+\lambda_{2}}{g_{3}(\tau)+ig_{5}(\tau)},1,0\right)^{\text{T}},\label{phi2}\\
|\phi_{3}(\tau)\rangle&=\eta^{(b)}_{3}\left(\frac{g_{1}(\tau)+\lambda_{3}}{g_{6}(\tau)+ig_{2}(\tau)},0,0,1\right)^{\text{T}},\label{phi3}\\
|\phi_{4}(\tau)\rangle&=\eta^{(b)}_{4}\left(\frac{g_{1}(\tau)+\lambda_{4}}{g_{6}(\tau)+ig_{2}(\tau)},0,0,1\right)^{\text{T}},\label{phi4}
\end{align}
\end{subequations}
where  $\eta_{j}^{(a)}=\sqrt{\left[g_{3}^{2}(\tau)+g_{5}^{2}(\tau)\right]/2\lambda_{j}\left[\lambda_{j}+g_{4}(\tau)\right]}$  and $\eta_{j}^{(b)}=\sqrt{\left[g_{6}^{2}(\tau)+g_{2}^{2}(\tau)\right]/2\lambda_{j}\left[\lambda_{j}+g_{1}(\tau)\right]}$ are normalization constants.

Now, let us consider the particular nonadiabatic evolution from the initial separable state $|\varphi(0)\rangle=|00\rangle$ to the maximally 
entangled state $|\varphi(1)\rangle=(|00\rangle+|11\rangle)/\sqrt{2}$  (target state), which corresponds to setting $j,k=0$ in Eq.~(\ref{QSE-ex}). 
For this purpose, we shall solve the set of differential equations \eqref{gequations}. In order to determine $\mathcal{I}(\tau)$, 
we fix the boundary conditions by defining $g_{1}(0)=\Gamma-\Delta^2/2\Gamma,$ $g_{2}(0)=0,$  $g_{6}(0)=\Delta$, and  
$g_{1}(1)=-\Delta^2/2\Gamma$, $g_{2}(1)=0,$ and  $g_{6}(1)=\Gamma$, where $\Delta$ and $\Gamma$ are arbitrary positive constants. 
From the eigenvalue Eq.~(\ref{ll4}), it follows that $\lambda_4$, $\Delta$, and $\Gamma$ are related by 
$\lambda_4 = \sqrt{\Gamma^2 + \Delta^4/(4\Gamma^2)}$. Moreover, we observe that, as the constant $\Delta$ tends to zero, 
the state $|\phi_{4}(\tau)\rangle$ tends to $|\varphi(0)\rangle$ in $\tau=0$ and tends to  $|\varphi(1)\rangle$ in $\tau=1$.  
Hence, $\Delta$ is closely connected with the fidelity of starting and ending in an eigenstate of $\mathcal{I}(\tau)$. We can then infer that the lower the value of $\Delta$ the closer the initial state and final state to the eigenstates of  
$\mathcal{I}(\tau)$. According to the chosen boundary conditions, we propose the following ansatz:
\begin{equation}
g_{1}(\tau)=\frac{\Gamma}{2}\left\{ 1+\cos\left[(2n-1)\pi \tau\right]\right\}-\frac{\Delta^2}{2\Gamma}, 
\end{equation}
where $n$ is a free integer parameter. As will be shown below, $n$ will be associated with the modulation of the transverse  
field, being directly related to the running time of the protocol.  
From Eq. \eqref{g1} we obtain
\begin{equation}
g_{2}(\tau)=-\frac{(2n-1)\Gamma}{4JT}\sin\left[(2n-1)\pi \tau\right].
\end{equation}
Applying the corresponding constraint \eqref{ll4}, we have
\begin{equation}
g_{6}(\tau)=\sqrt{\lambda_{4}^{2}-g_{1}^{2}(\tau)-g_{2}^{2}(\tau)}.
\end{equation}
Finally, employing the previous results, we can obtain the function  $f(\tau)$ from the differential equation \eqref{g6}; it reads
\begin{equation}
f(\tau)=\frac{2\pi TJg_{1}+\dot{g_{2}}}{4 T\sqrt{\lambda_{4}^2-g_{1}^2-g_{2}^2}} . 
\end{equation}
\begin{figure}[h]
\hspace*{-2 mm}
\includegraphics[scale=0.3]{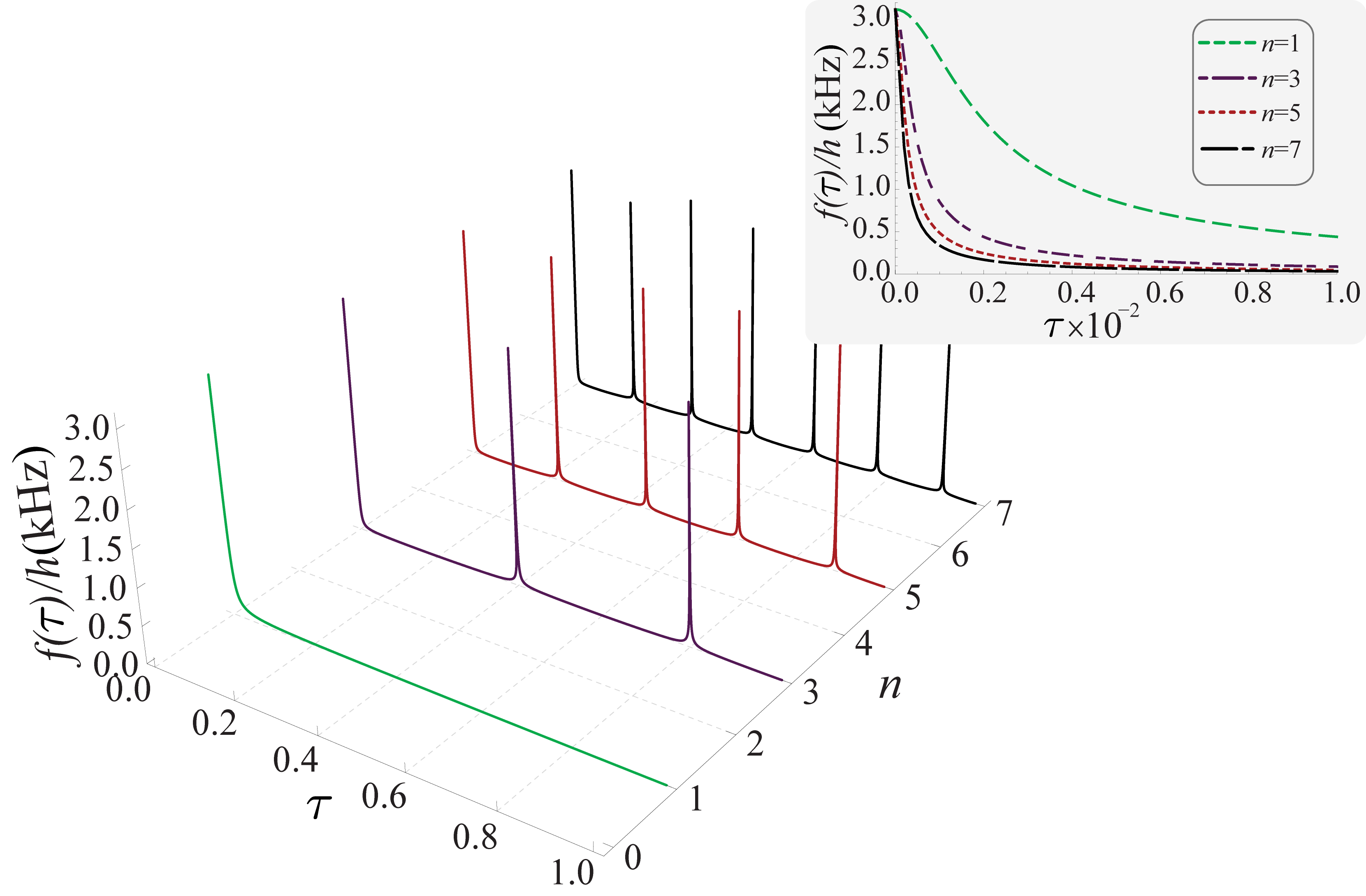}
\caption{(Color online) Transverse field quenches as given by the function $f(\tau)$ for $n=1,3,5,7$. We have taken 
$\varepsilon=(\pi/2)\times10^{-2}$,  $\omega=1$, $J=1$, $\Delta^2=10^{-5}$, and  $\Gamma=1$.}
\label{rapidquench}
\end{figure}%
Notice that  if $\Delta$ is very small,  $f(\tau)$ is very large at $\tau=0$.  The function $f(\tau)$ is an external control source and 
depends on the experimental capabilities. In the case of an NMR setup this parameter is controlled by a transverse magnetic field, 
so that  the allowed values of $f(\tau)$ will be limited by the resources available in the laboratory. 
The coefficients $g_3(\tau)$, $g_4(\tau)$, and $g_5(\tau)$ do not directly contribute to the evolution dynamics of the system. 
Their general solution is given by
\begin{eqnarray}
g_{3}(\tau)&=&g_{3}(0),\\
g_{4}(\tau)&=&g_4(0)\cos\left[2\pi JT\tau\right]+g_5(0)\sin\left[2\pi JT\tau\right],\\
g_{5}(\tau)&=&g_5(0)\cos\left[2\pi JT\tau\right]-g_4(0)\sin\left[2\pi JT\tau\right],
\end{eqnarray}

\begin{figure}[h]
\hspace*{-6 mm}
\includegraphics[scale=0.32]{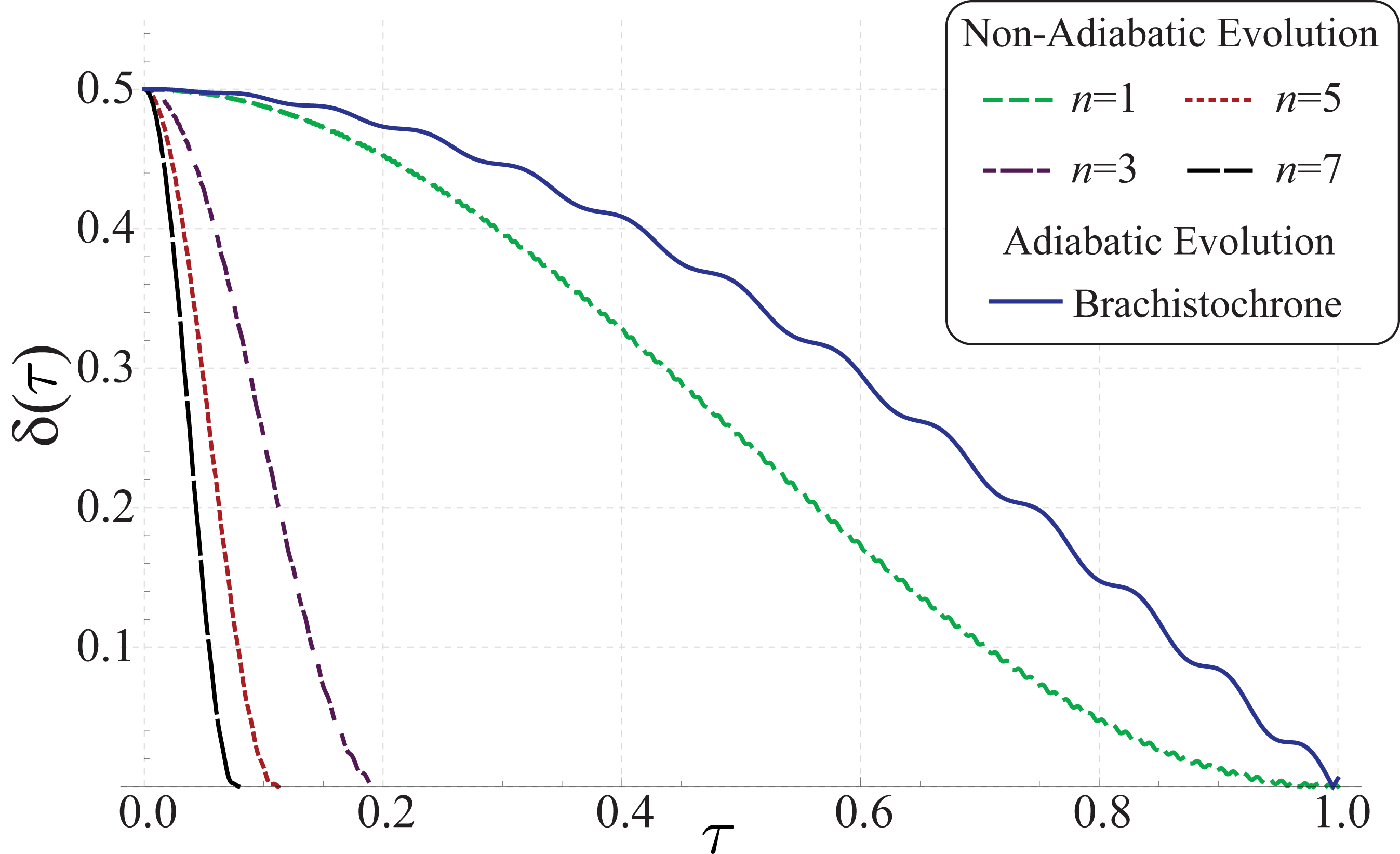}
\caption{(Color online) Trace distance for the nonadiabatic evolution with different values of $n$  and for the adiabatic evolution.  
We have taken $\varepsilon=(\pi/2)\times10^{-2}$,  $\omega=1$, $J=1$, $\Delta^2=10^{-5}$, and $\Gamma=1$, and we have parametrized the time by the total time of the adiabatic evolution (in which we have $\tau = 1$).}
\label{fig-tracedistance}
\end{figure}

To investigate the performance of this nonadiabatic route,  we solve numerically the Hamiltonian dynamics for both adiabatic and nonadiabatic 
evolutions.  The results are then shown in Fig.~\ref{rapidquench}, which displays the fast quenches given by the modulation function 
$f(\tau)$. We can observe that the function starts with a high intensity value ($3.14$ kHz) and it decays rapidly to zero for different 
choices of $n$. This intensity value is consistent with current NMR experimental setups. The details of the field quench can be 
observed in the inset of Fig.~\ref{rapidquench}. This time-modulated  field induces a dynamics that does not satisfy the adiabatic 
condition. It is worth mentioning that higher values of $n$ lead to a faster dynamics for the system.  In Fig.~\ref{fig-tracedistance}, 
we  show the trace distance given by Eq.~(\ref{tracedistance}) for the adiabatic QSE (continuous line) and for nonadiabatic 
QSE (noncontinuous lines), with the normalization of time chosen in such a way that the target state is reached at $\tau=1$ in the adiabatic dynamics. 
From this plot, we observe that the trace distance vanishes rapidly for the nonadiabatic QSE, more rapidly as $n$ is increased.  

\section{Energetic cost for the dynamical evolution}
\label{workdistribution}
We can evaluate the energetic cost of both the adiabatic and nonadiabatic QSE in terms of the work performed on or by the system 
due to the field quench. In the quantum context, work acquires a meaning in a statistical average, defined as the mean value of a distribution containing many possible paths \cite{talkner,batalhao}, i.e.,  $\langle W\rangle = \int{W P(W)dW}$. The work probability 
distribution $P(W)$ is associated with the evolution generated by the quench on the system. Indeed, such a distribution has 
been observed at a quantum level, by employing an interferometric approach \cite{dorner,mazzola}, in a recent NMR experiment \cite{batalhao}. 
In the present case, the evolution of the system is driven by the field modulation function $f(\tau)$ for $\tau=[0,\tau_\textrm{f}]$, 
where $\tau_\textrm{f}$ is the final parametrized time necessary to prepare the desired state with a given accuracy, with 
$\delta(\tau) \ll 1$. The Hamiltonians at the initial and final configurations can be written through their spectral decompositions 
as  $H(0)=\sum_n \epsilon_n \ket{n}\bra{n}$ and $H(\tau_\textrm{f})=\sum_{m'} \epsilon_{m'} \ket{m'}\bra{m'}$. The work distribution 
for the quenched dynamics of a quantum system is given by \cite{talkner}
\begin{equation}
P(W):=\sum_{n,m'}p_n p_{m'|n}\delta(W-\Delta \epsilon_{m',n}),
\end{equation}
where $p_n$ is the probability to find the system in state $|n\rangle$ (with energy $\epsilon_{n}$) at $\tau=0$ , $ p_{m'|n}=|\langle m'| U(0,\tau_\textrm{f}) |n\rangle|^2 $ is the conditional probability to drive the system to $|m'\rangle$ (with energy $\epsilon_{m'}$) at the 
end of the quench protocol given the initial state $|n\rangle$, $U(0,\tau_f)$ is the time-ordered evolution operator, and $\Delta \epsilon_{m',n}=\epsilon_{m'}-\epsilon_n$ is the energy difference between the initial and final Hamiltonian spectra.  Using this distribution, the 
average work can be expressed as 
\begin{equation}
\langle W\rangle=\sum_{n,m'}p_n p_{m'|n} \Delta\epsilon_{m',n}.
\label{work}
\end{equation}
According to Eq. (\ref{work}), it is always possible to find an adiabatic interpolation Hamiltonian such that $\langle W \rangle=0$, 
where the system evolves to its instantaneous ground state at the end of the 
evolution, implying $\epsilon_{m',n}=0$ as in the example presented in Sec.~\ref{aqse}.  In this sense, the adiabatic strategy for QSE has 
no energetic cost.  Concerning the non-adiabatic strategy, the system evolves into a fixed instantaneous eigenstate of 
the dynamical invariant. This may encompass transitions in the Hamiltonian energy spectrum. Therefore the energetic investment 
will be typically higher in the nonadiabatic case. In turn, the profit of this investment is the smaller evolution period, as illustrated 
in Fig.~\ref{fig-tracedistance}.
\begin{figure}[h]
\hspace*{-6 mm}
\includegraphics[scale=0.33]{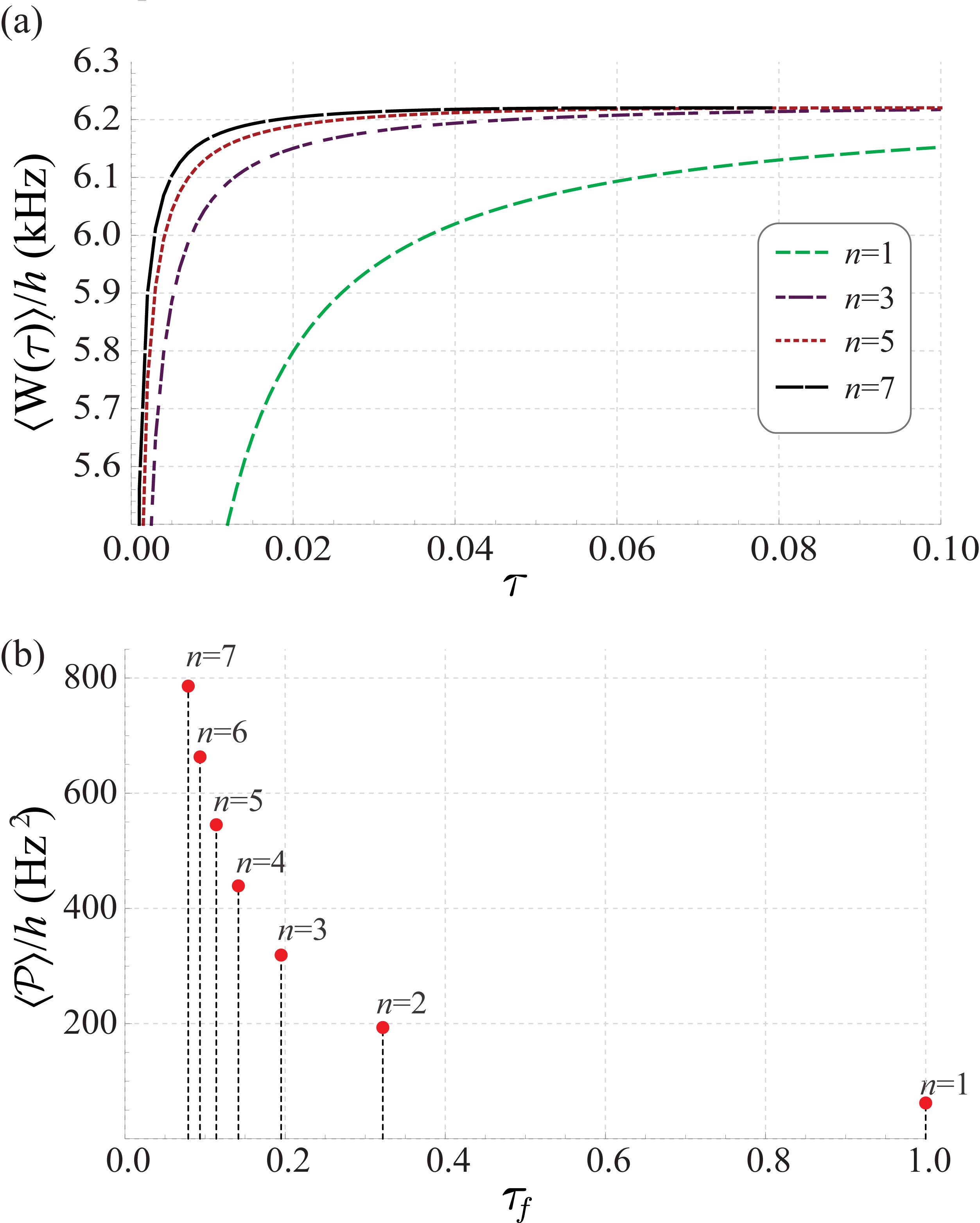}
\caption{(Color online)  Energetic cost of nonadiabatic QSE. (a) The average work per quench as a function of the time evolution. 
(b) Average power for different quenched evolutions (fast and slow evolutions). We have taken  
$\varepsilon=(\pi/2)\times10^{-2}$,  $\omega=1$, $J=1$, $\Delta^2=10^{-5}$, and $\Gamma=1$, and we have parametrized the time by the total time of the adiabatic evolution (in which we have $\tau=1$). The average work per quench is the work employed in each field quench displayed in Fig. 1. We note that, for $n \geq 2$, the evolution is cyclic (as also depicted in Fig. 1).  Here, we are assuming that when the target state is reached, the control field is switched off.}
\label{fig-work}
\end{figure}%

The average work invested [according to Eq. \eqref{work}] in the nonadiabatic QSE is presented in Fig.~\ref{fig-work}(a).  
The mean invested work has the same value for different quenches (rapid or slow).  What really distinguishes the quench 
dynamics is the rate at which the average work is performed. In other words, we will be interested in the average power 
[as displayed by Fig. \ref{fig-work}(b)], which can be defined as $\langle \mathcal{ P}\rangle=\frac{\langle W\rangle}{\Delta t}$, 
where the work $\langle W\rangle$ is to be considered as performed during a period of time of duration $\Delta t$. 
Using the normalized time, we can rewrite the time interval as $\Delta t = T_{ad}\tau_{\textrm{f}}$.  In fact, the average 
power, defined as the rate at which energy is introduced by the driven field, can be used as a figure of merit to quantify the energy 
investment required to perform a faster evolution. As can be observed in Fig.~\ref{fig-work}, a faster quench implies a large power consumption. Hence, the state preparation task can be accomplished in a long time by a low-powered system or can be accomplished in a short time 
by a high-powered system. In the end, the running time limit for the nonadiabatic protocol proposed  here is associated with the 
maximum power of the driven field modulator. 

\section{Conclusion}
We presented a method, employing fast quenches obtained via dynamic invariants, which enables us to tailor a shortcut to the 
adiabatic QSE. Such a strategy has been shown to be applicable to an experimentally realizable scenario of NMR systems. The 
energetic cost of this shortcut has  also been investigated, with the limit of such nonadiabatic performance being associated 
with the average power of the driven system. The generalization of our fast-quench approach for the case of open systems is 
left as a future challenge. We believe this generalization may reveal in a more explicit (and realistic) picture the advantage of 
the nonadiabatic QSE, since decoherence is expected to reduce the accuracy in the preparation of the target state as time 
increases. Moreover, the experimental realization of the protocol (e.g., in an NMR setup) remains for a future work, along with the observation 
of the energetic cost for the shortcut in a controllable scenario.
\begin{acknowledgments}
We acknowledge financial support from the Brazilian agencies CNPq, CAPES, FAPERJ, and FAPESP. This work was performed as part of the 
Brazilian National Institute of Science and Technology for Quantum Information (INCT-IQ).
\end{acknowledgments}


\end{document}